\documentclass[12pt]{iopart}

\usepackage[dvips]{graphicx}

\begin{document}
\letter{Additional selection rule for some emission, photoexcitation and Auger spectra}

\author{
Andrius Bernotas and Romualdas Karazija
}

\address {State Institute of Theoretical Physics and Astronomy,
A.~Go\v{s}tauto 12, 2600, Vilnius, Lithuania}

\eads{\mailto{bernotas@itpa.lt}, \mailto{karazija@itpa.lt}}

\begin{abstract}
	Concentration of the strongest transitions on the high-energy side of some emission, photoexcitation and Auger spectra of atoms is related with the existence of the additional selection rule for the number of vacancy-electron pairs.
\end{abstract}

\vspace{28pt plus 10pt minus 18pt}
\noindent{\small\rm Published in: {\it \jpb} {\bf V34} No22 L741\par}

\noindent{\small\rm Online at: {\it http://stacks.iop.org/JPhysB/34/} \par}


\maketitle

	The concentration of the oscillator strength near the high-energy side of the transition array was first established for the $4{\rm d}^94{\rm f}^N \rightarrow 4{\rm d}^{10}4{\rm f}^{N-1}$ photoexcitation spectra from lanthanides \cite{Sugar}. When the $4{\rm f}$ orbital is collapsed the most intensive lines of these transitions form the so-called giant resonances \cite{GiantRes}. The concentration of the strongest transition lines manifests itself also from some characteristic emission spectra. This regularity was noticed in a survey of experimental analysis of $3{\rm p}^53{\rm d}^N \rightarrow 3{\rm p}^63{\rm d}^{N-1}$ transitions \cite{Ryabtsev}. Such a tendency was also demonstrated by the calculations of complex spectra for various Sm ions \cite{O'Sullivan}. This effect explains the appearance of narrow continuum band in the emission of laser produced plasma from rare-earth elements \cite{Costello}. The shift to the high-energy side of some complex Auger spectra corresponding to transitions from the initial $nl_1^{N_1}nl_2^{N_2}$ configuration was predicted and illustrated by calculations for lanthanide atoms in \cite{JKK97}.

	Such an effect takes place for the configurations with two open shells with the same principal quantum number, when the Coulomb exchange interaction mainly determines the energy level spectrum. This interaction forms the upper and the lower groups of levels,  with very different abilities to participate in the transitions. Due to the relation between the position of level and the transition amplitude from this level \cite{Sugar, KK85, KK91}, the transitions from mainly the upper group of levels of initial configuration manifest themselves in the radiative or Auger spectra.

	This enhancement of some transitions and near-forbiddenness of the other transitions indicates the existence of an additional selection rule. The aim of this work is the formulation of such a rule.

	In the case of the configuration with one vacancy $nl^{4l+1}n(l+1)^{N_2}$ it was suggested \cite{Sugar} (and generalised for any configuration with two open shells $nl^{N_1}n(l+1)^{N_2}$ \cite{KK85}) to classify the energy levels in a way which preserves $nl^{4l+1}n(l+1)$ parentage. Such basis can be obtained by diagonalizing the matrix of the main Coulomb exchange coefficient $g_1$. The algebraic way to construct this basis by using a special operator $A^{(10)}$ was proposed in \cite{22EGAS}. Operator  $A^{(10)}$ is defined by:

\begin{equation}
\label{eq1}
A^{(10)}={\left[{\tilde b}^{(l_1s)\dagger} \times a^{(l_2s)\dagger} \right]}^{(10)}
\end{equation}
where $a^{(l_2s)\dagger}$ is an electron creation operator with an orbital rank $l_2$ and spin rank $s$, $b^{(l_1s)\dagger}$ is a vacancy creation operator, equal to the electron annihilation operator $a^{(l_1s)}$. $a^{(l_1s)}$ only becomes the irreducible tensor when multiplied by a phase factor, thus in (\ref{eq1}) the following operator is used:

\begin{equation}
\widetilde{b}_{m_{1}\mu_{1}}^{(l_{1}s)\dagger }
= (-1)^{l_1+s-m_1-\mu_1}b_{-m_1-\mu_1}^{(l_1s)\dagger } 
= (-1)^{l_1+s-m_1-\mu_1}a_{-m_1-\mu_1}^{(l_1s)}
~.
\end{equation}

	We shall consider the configurations $nl^{N_1}n(l+1)^{N_2}$ with two neighbouring shells, they are designated $l_1^{N_1}l_2^{N_2}$. While acting upon the wavefunction of configuration $l_1^{N_1+1}l_2^{N_2-1}$ the operator $A^{(10)}$ creates a vacancy in the $l_1^{N_1+1}$ shell and an electron in the $l_2^{N_2-1}$ shell. Thus the wavefunction of configuration $l_1^{N_1}l_2^{N_2}$ is obtained. Due to the ranks of operator $A^{(10)}$ in the orbital ($=1$) and spin ($=0$) spaces the vacancy-electron pair $l_1^{-1}l_2$ is coupled to a term $^{1}{\rm P}$.

	We introduce a new quantum number $p$ equal to the number of such vacancy-electron pairs. The configuration $l_1^{4l+2}l_2^{N_2}$ has no vacancies in $l_1^{4l_1+2}$ shell, so $p=0$. After acting by operators $A^{(10)}$ repeatedly, various states of configurations $l_1^{N_1}l_2^{N_2'}$ ($N_1+N_2' = 4l_1+2+N_2$) with different numbers $p$ can be obtained. To be exact, the complete basis for a given configuration may be built with the aid of operators $A^{(K\kappa)}$ having general ranks $K$ and $\kappa$:
\begin{eqnarray}
\label{eq3}
\fl \left| l^{N_{1}}_{1}l^{N_{2}}_{2}\gamma pLSM_{L}M_{S}\right\rangle  = \frac{1}{\sqrt{N_{1}N_{2}}} \sum_{K\kappa,\gamma'p'L'S'} \left[ A^{(K\kappa)}\times  \left| l^{N_{1}+1}_{1}l^{N_{2}-1}_{2}\gamma' p' L'S' \right\rangle \right]^{(LS)}_{M_{L}M_{S}} \nonumber \\
\lo{\times} \left( l^{N_{1}}_{1}l^{N_{2}}_{2}\gamma pLS\right. \left\Vert l^{N_{1}+1}_{1}l^{N_{2}-1}_{2}\gamma' p' L'S, l_1^{-1}l_2~K\kappa\right)~
\end{eqnarray}
where $p=p'+1$ if the fractional parentage coefficient (the last multiplier on the right-hand side) for $K\kappa=10$ is non-zero, and remains $p=p'$ otherwise. The labelling of those basis functions would generally require some other quantum numbers $\gamma$ of a two-shell system besides the number of $^{1}{\rm P}$ pairs $p$ and the resultant $LS$. However, only $p$ is important for this paper. This new basis was named the hole-particle (HP) basis \cite{22EGAS}. It can be chosen in such a way that for the configuration $l_1^{4l+2}l_2^{N_2}$ its functions coincide with the functions of a single open shell of the usual basis. The values of these coefficients can be calculated using the recurrency relations similar to those derived for the two-electron coefficients \cite{ParticleHole,Thesis,CFP}.

	In the second quantization representation the matrix element of Coulomb exchange interaction between the two open shells is:
\begin{eqnarray}
\label{eq4'}
\fl \left\langle l^{N_{1}}_{1}l^{N_{2}}_{2}\gamma pLSM_{L}M_{S}\right|H^{e}_{ex}\left| l^{N_{1}}_{1}l^{N_{2}}_{2}\gamma' p'LSM_{L}M_{S}\right\rangle  \equiv \sum_{k} g_{k}(l_{1}^{N_{1}}l_{2}^{N_{2}}\gamma p\gamma 'p'LS)~G^{k}(l_{1},l_{2}) \nonumber \\
\fl = \sum_{k} \left\langle l_{1}\left\Vert C^{(1)}\right\Vert l_{2}\right\rangle ^{2} \\
\fl \times \left\langle l^{N_{1}}_{1}l^{N_{2}}_{2}\gamma pLSM_{L}M_{S}\right| \frac{2}{2k+1}\left(A^{(k0)}\cdot \widetilde{A}^{(k0)}\right)  - \frac{N_{2}}{2l_{2}+1} \left| l^{N_{1}}_{1}l^{N_{2}}_{2}\gamma' p'LSM_{L}M_{S}\right\rangle  G^{k}(l_{1},l_{2}) \nonumber
\end{eqnarray}
with $G^{k}(l_{1},l_{2})$ being the Slater radial integrals. The expression for the angular coefficient $g_1$ at the main (dipole) part of this interaction is easily obtained from the above by putting a projector onto complete basis $\left| l^{N_{1}+1}_{1}l^{N_{2}-1}_{2}\gamma'' p''L''S''M''_{L}M''_{S}\right\rangle$ in between the two $A^{(10)}$-type operators (creation and annihilation of a $^{1}{\rm P}$-coupled pair)\cite{KK91,CFP}:
\begin{eqnarray}
\label{eq4}
\fl g_{1}(l_{1}^{N_{1}}l_{2}^{N_{2}}\gamma p\gamma 'p'LS) = \delta (p,p')\left\langle l_{1}\left\Vert C^{(1)}\right\Vert l_{2}\right\rangle ^{2}  \nonumber \\
\fl{\times} \left\{ 2(4l_{1}+2-N_{1})N_{2}\sum_{\gamma ''L''}\right.\left( l^{N_{1}+1}_{1}l^{N_{2}-1}_{2}\gamma'' p-1~L''S, l_1^{-1}l_2~^{1}{\rm P}\right\Vert \left. l^{N_{1}}_{1}l^{N_{2}}_{2}\gamma pLS\right) \\
\lo{\times} \left( l^{N_{1}+1}_{1}l^{N_{2}-1}_{2}\gamma'' p-1~L''S, l_1^{-1}l_2~^{1}{\rm P}\right\Vert \left. l^{N_{1}}_{1}l^{N_{2}}_{2}\gamma' pLS\right) -\left. \delta (\gamma,\gamma') \frac{N_{2}}{2l_2+1} \right\} ~.\nonumber
\end{eqnarray}
The first term-dependent part of this coefficient would remain non-diagonal with respect to the quantum numbers $\gamma, \gamma'$ unless we demand explicitly that all the fractional parentage coefficients but one involved in (\ref{eq4}) are equal to $0$ for the given $\gamma p LS$ term. This is done by exploiting the freedom of choice of fractional parentage coefficients for the repeating terms with same resultant $LS$. Then the first, positive, part of $g_1$ contributes only to those terms of configuration $l_1^{N_1}l_2^{N_2}$ which are the daughters of $l_1^{N_1+1}l_2^{N_2-1}$ terms (additionally characterized by $\gamma''~p-1$) and a $l_1^{-1}l_2$ $^1{\rm P}$ pair, and it vanishes for the other terms that do not satisfy the condition of having an additional vacancy-electron pair $^1{\rm P}$ (in this case the coefficient $g_1$ takes the negative constant value determined by the second part of $g_1$ in (\ref{eq4})).

	The selection rule for the dipole transition amplitude with respect to the quantum number $p$ follows from the simple relation between the operator $A^{(10)}$ and the radiative dipole transition operator $D^{(1)}$ in the second quantization representation for the considered transitions $l_1^{N_1}l_2^{N_2} \rightarrow l_1^{N_1+1}l_2^{N_2-1}$ :
\begin{equation}
\label{eq5}
D^{(1)}_{~q} = \sqrt{\frac{2}{3}}A^{(10)}_{~q0} 
\langle l_{1}\Vert C^{(1)}\Vert l_{2}\rangle
\langle n_1l_1\left.\right| r \left|\right. n_2l_{2}\rangle
~.
\end{equation}
One obtains:
\begin{eqnarray}
\label{eq6}
\fl \left\langle l_1^{N_1}l_2^{N_2} \gamma p LS M_L M_S \left| D^{(1)}_{~q} \right| l_1^{N_1+1}l_2^{N_2-1} \gamma' p' L'S M_{L'} M_S\right\rangle = \delta(p,p'+1)\sqrt{\frac{2N_1N_2}{3}} \nonumber\\
\lo{\times} {\left[ \begin{array}{ccc} 1 & L' & L \\ q & M_{L'} & M_{L} \end{array}\right]} \left( l^{N_{1}+1}_{1}l^{N_{2}-1}_{2}\gamma' p'~L'S, l_1^{-1}l_2~^{1}{\rm P}\right\Vert \left. l^{N_{1}}_{1}l^{N_{2}}_{2}\gamma pLS\right) \\
\lo{\times} \langle l_{1}\Vert C^{(1)}\Vert l_{2}\rangle \langle n_1l_1\left.\right| r \left|\right. n_2l_{2}\rangle \nonumber
\end{eqnarray}
where the quantity in square brackets is the Clebsch-Gordan coefficient. Thus in the pure HP coupling only the transitions obeying the rule $\Delta p=p-p'=1$ are possible. The idea of such a selection rule was suggested in \cite{Sums}. There were also attempts to explain the preference of some transitions using the usual basis: it was indicated in \cite{O'Sullivan,Karosiene} that the dipole transitions from $4{\rm d}^{10}4{\rm f}^N$ mainly went to the states of $4{\rm d}^94{\rm f}~^1{\rm P}~4{\rm f}^N$ (however, all electrons of the same shell must be treated equivalently in this basis).

	The Auger transition operator corresponding to the transitions
\begin{eqnarray}
\label{eq7}
K_0nl^{N_1}n(l+1)^{N_2} \longrightarrow K_0nl^{N_1+1}n(l+1)^{N_2-2}\epsilon l' ~,\nonumber\\
K_0nl^{N_1}n(l+1)^{N_2}n_3l_3^{N_3} \longrightarrow K_0nl^{N_1+1}n(l+1)^{N_2-1}n_3l_3^{N_3-1}\epsilon l'
\end{eqnarray}
can be expressed by a scalar product of operator $A^{(10)}$ (\ref{eq1}) and the other similar operator creating a vacancy in $n(l+1)^{N_2}$ or $n_3l_3^{N_3}$ shell and a free electron \cite{JKK97}. Consequently, in the HP basis the Auger transition amplitude is presented as the product of two matrix elements; for the first of them the selection rule with respect to the $p$ quantum number also applies.

	The selection rule for the number of pairs will manifest itself in the emission spectra if the HP basis describes the involved configurations well. The comparison of the wavefunction expansions in the HP and in the usual basis was performed for the Rb~{\small IV} $4{\rm p}^34{\rm d}$ and Ba~{\small XII} $4{\rm d}^84{\rm f}$ \cite{KK91} configurations as well as for K~{\small I} $3{\rm p}^53{\rm d}^2$ and Rb~{\small I} $4{\rm p}^54{\rm d}^2$ isoelectronic sequences \cite{Thesis}. In all these cases the leading weights become larger for the majority of levels and even close to 1 for several levels (especially from the upper group) in an HP basis. In table \ref{weights} the results for the upper terms which mainly take part in the radiative transitions are presented for Co~{\small IX} $3{\rm p}^53{\rm d}^2$ and Co~{\small VIII} $3{\rm p}^53{\rm d}^3$. High purities of the wave functions confirm the conjecture that this basis suits well for the description of considered spectra. Some reasons for the preciseness of the HP basis may be pointed out: (a) the main part of the interaction dominating in such configuration only has diagonal matrix elements; (b) in the HP basis some matrix elements connecting such configuration with the other configuration of the same complex vanish \cite{KK91} (since the corresponding operator, similarly as the Auger transition operator, can be expressed by the operator $A^{(10)}$ (\ref{eq1})). 

\begin{table}
\caption{\label{weights}Purity (leading weight) for the terms of upper group of $3p^53d^N$ from Co~{\small IX} ($N=2$) and Co~{\small VIII} ($N=3$) in the $LS$ and HP bases. The $3d^N~$ terms, with $v$ if necessary, are given in brackets for complete labelling in the usual basis. Term energies decrease top to bottom.}
\begin{tabular*}{\textwidth}{@{}l*{15}{@{\extracolsep{0pt plus12pt}}l}}
\br
Configuration& Term, & \multicolumn{2}{c}{Leading weight} \\
&$(^{2S'+1}L')^{2S+1}L$&usual $LS$ basis&HP basis\\
\mr

$3{\rm p}^{5}3{\rm d}^2$&$(^{3}{\rm F})^{2}{\rm D}$ & 0.7137&0.9996\\
&$(^{3}{\rm P})^{2}{\rm P}$ & 0.7452&0.9990 \\
&$(^{3}{\rm F})^{2}{\rm F}$ & 0.5132&0.9815 \\
& & & \\
$3{\rm p}^{5}3{\rm d}^3$&$(_{1}^{2}{\rm D})^{1}{\rm P}$ & 0.7097& 0.9476\\
&$(^{4}{\rm P})^{3}{\rm S}$ & 0.7884& 0.9992\\
&$(^{2}{\rm F})^{1}{\rm F}$ & 0.3710& 0.9977\\
&$(^{2}{\rm H})^{1}{\rm G}$ & 0.5286& 0.9973\\
&$(^{4}{\rm F})^{3}{\rm D}$ & 0.5870& 0.9982\\
&$(^{4}{\rm F})^{3}{\rm F}$ & 0.6714& 0.9948\\
&$(^{4}{\rm P})^{3}{\rm P}$ & 0.4899& 0.9859\\
&$(^{2}{\rm F})^{1}{\rm D}$ & 0.3435& 0.9914\\
&$(^{2}{\rm H})^{1}{\rm H}$ & 0.9991& 0.9386\\ 
&$(^{2}{\rm H})^{3}{\rm G}$ & 0.6240& 0.9073\\
\br
\end{tabular*}
\end{table}

\begin{figure}
\begin{center}
\includegraphics[angle=180, width=0.9\textwidth]{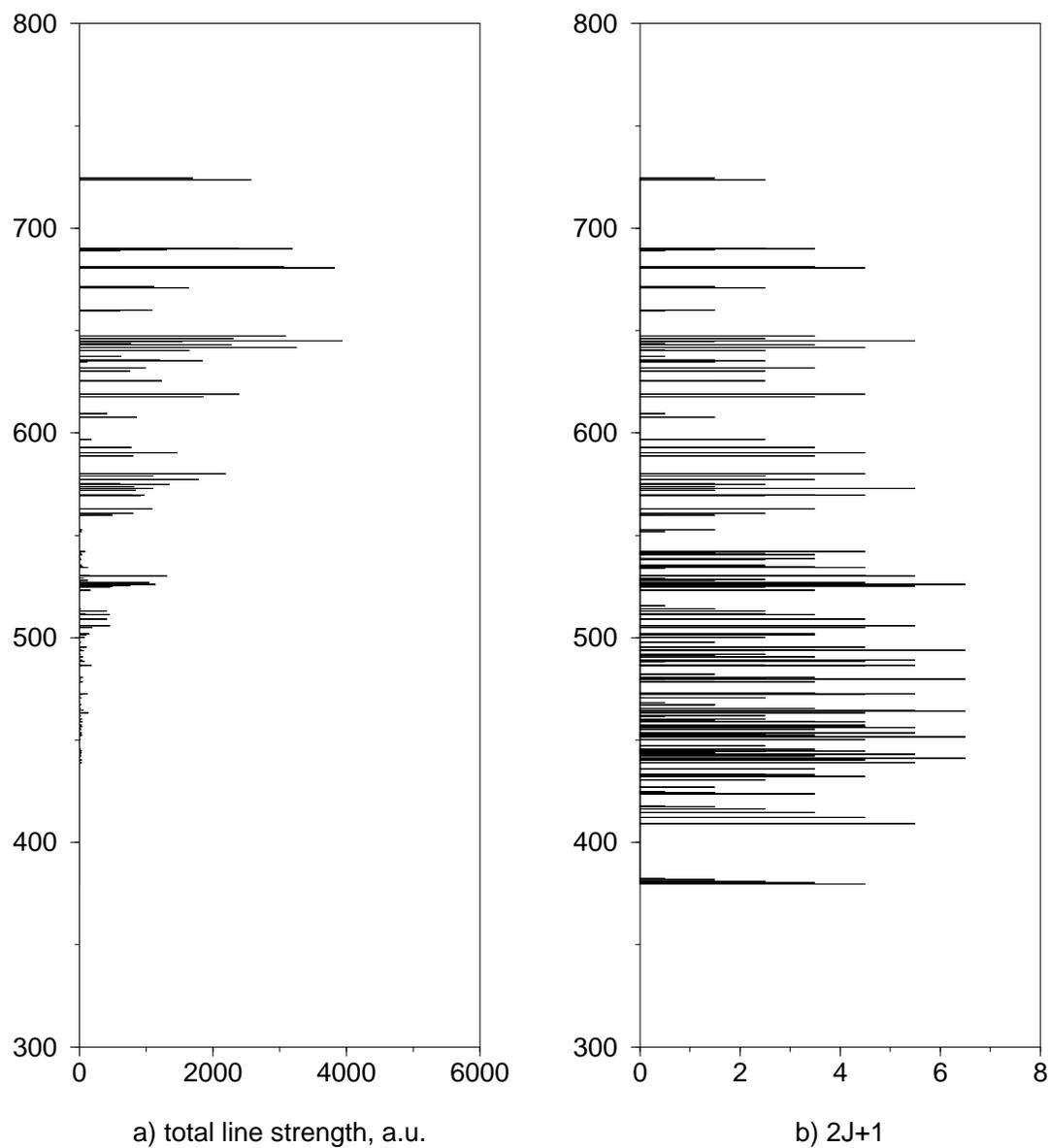}
\end{center}
\caption{\label{tstrength} Distribution of the total line strengths for transitions from the levels of Co~{\small VII} $3{\rm p}^53{\rm d}^4$ to all levels of $3{\rm p}^63{\rm d}^3$: $(a)$ results of calculation in the single configuration intermediate coupling approximation; $(b)$ total line strength taken to be proportional to the statistical weight of the initial level (all states participate equally in the transitions). On the vertical axis the binding energies of levels of $3{\rm p}^53{\rm d}^4$ are indicated, in $1000cm^{-1}$.}
\end{figure}

	The calculations in intermediate coupling give small intensities of lines that are forbidden by the indicated selection rule from the lower group of levels of excited configuration. Their insignificant participation in the formation of spectrum is illustrated by figure \ref{tstrength}, in which the total line strenghts from the given levels of initial configuration to all levels of final configuration are presented for Co~{\small VII} $3{\rm p}^53{\rm d}^4 \rightarrow 3{\rm p}^63{\rm d}^3$. The distribution of total line strengths differs essentially if all states are supposed to take equal part in the transitions (all levels participate in proportion to their statistical weights). The suppression of transitions from lower levels is rather strict.

	The more significant violations of the rule $\Delta p = 1$ may be caused by correlation effects. However, as we have indicated above, some interconfiguration matrix elements in HP basis vanish, and it also concerns the fairly important mixing of configuration $nl^{4l+1}n(l+1)^N$ with $nl^{4l+2}n(l+1)^{N-2}n(l+2)$.

\begin{figure}
\begin{center}
\includegraphics[width=0.9\textwidth]{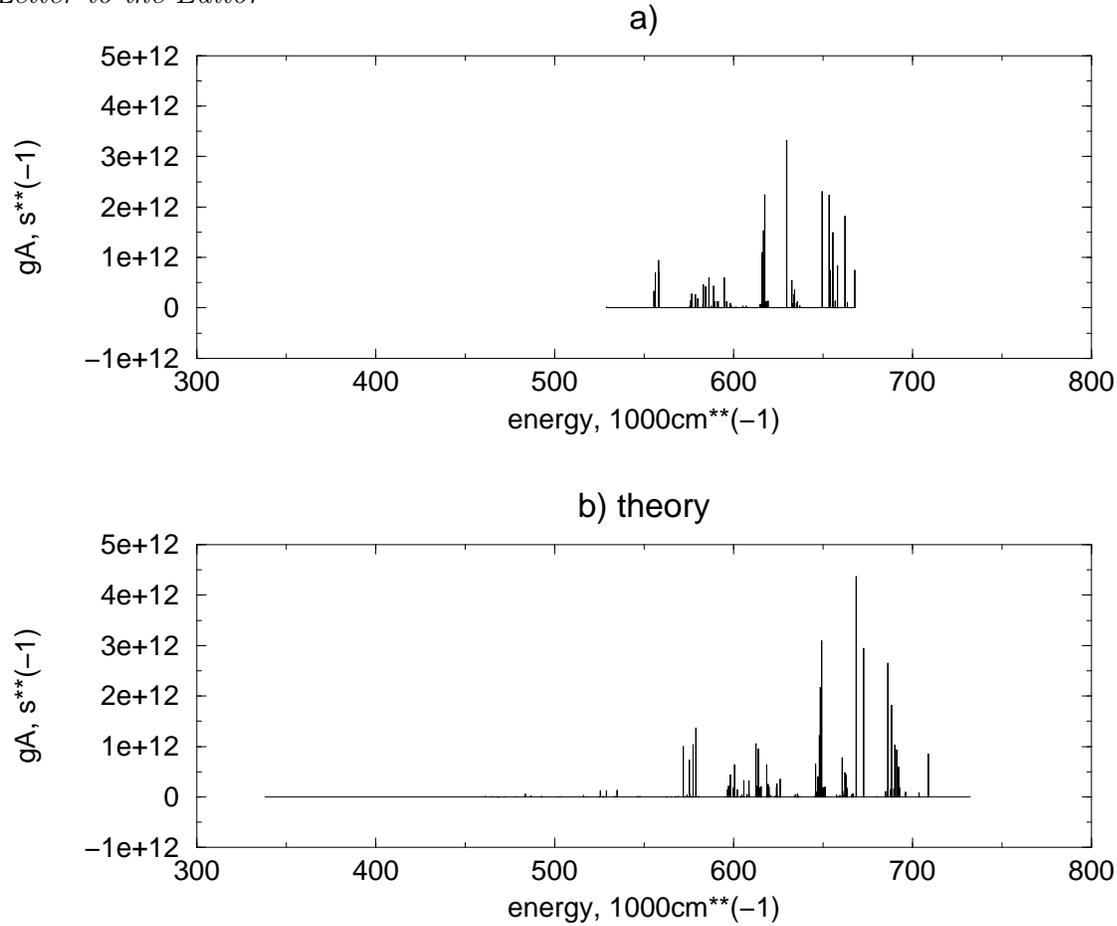}
\end{center}
\caption{\label{ExpTh} Spectrum of radiative transitions for Co~{\small VIII} $3{\rm p}^53{\rm d}^3 \rightarrow 3{\rm p}^63{\rm d}^2$: $(a)$ reference \cite{Shirai}; $(b)$ results of calculation in single configuration approximation. $A$ is the transition probability, $g$ is the statistical weight of the initial level. The bottom horizontal line in the (a) graph indicates the interval of available data and in the (b) graph the total interval of transition energies.}
\end{figure}

	Therefore, the suppression of low energy transitions and the enhancement of the other transitions clearly manifest themselves in various experimental spectra. For example, in figure \ref{ExpTh} the results of calculation in single configuration approximation of the radiative transition probabilities $A(3{\rm p}^53{\rm d}^3 \gamma J - 3{\rm p}^63{\rm d}^2 \gamma' J')$ multiplied by the statistical weights $g=2J+1$ are compared for Co~{\small VIII} with the data tabulated in \cite{Shirai}. Both spectra have the same qualitative features. All strong lines are located on the high energy side of the spectrum and only a few weak lines are seen in the middle of the spectrum. Experimentally only intensive lines from the upper group of levels are registered.

	Let us consider briefly how the concentration of intensive lines on the high-energy side of the spectrum and the suppression of other lines depend on the number of electrons in the outer open shell. For this purpose we can use the shift of the average energy of emission spectrum (~$\overline{E}(K-K')$~) with respect to the difference of average energies of initial (~$\overline{E}(K)$~) and final (~$\overline{E}(K')$~) configurations:
\begin{equation}
\label{eq8}
\delta \overline{E}(K - K') = \overline{E}(K - K') - 
\left[ \overline{E}(K) - \overline{E}(K')\right]~.
\end{equation}
The algebraic expression for this quantity, averaged with the weight equal to the square of the dipole transition amplitude, was obtained in \cite{Bauche}.

\begin{figure}
\begin{center}
\includegraphics[angle=-90, width=0.9\textwidth]{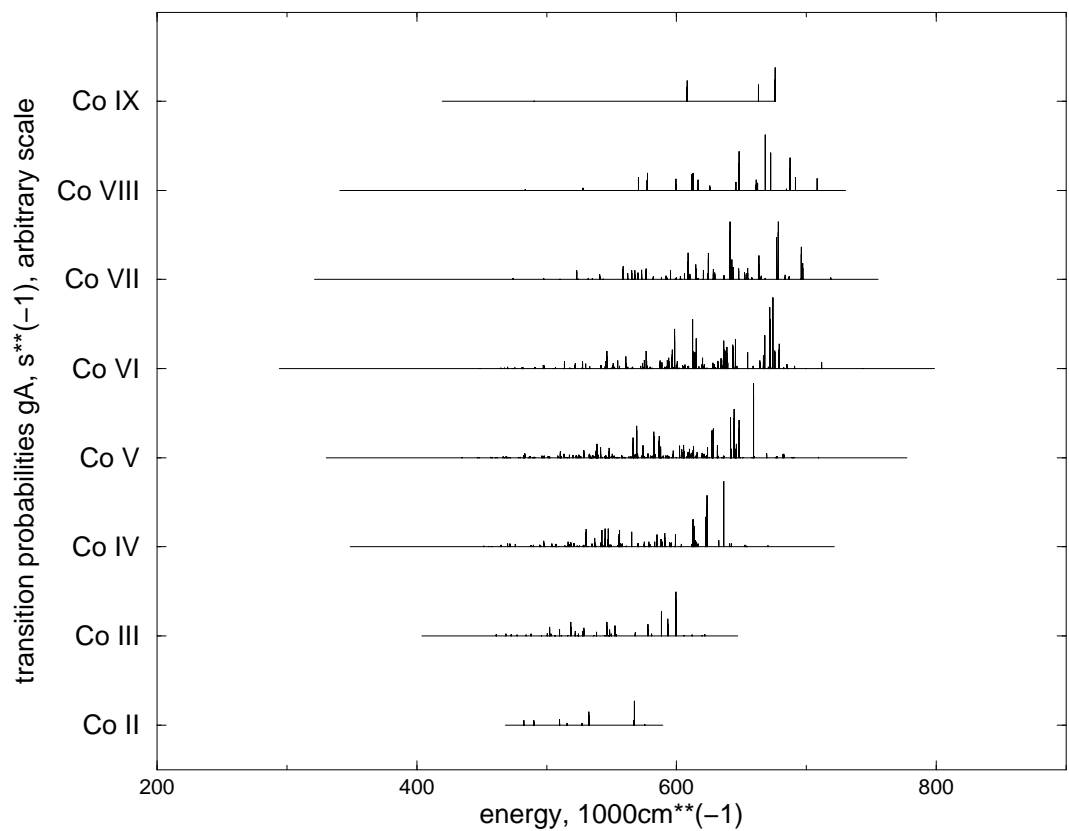}
\end{center}
\caption{\label{allCo} Variation of the spectrum of radiative transitions in the isonuclear sequence $3{\rm p}^53{\rm d}^N \rightarrow 3{\rm p}^63{\rm d}^{N-1}$ of Co. The horizontal sections of graphs indicate the intervals of possible transition energies.}
\end{figure}

	For the transitions between two neighbouring shells with the same principal quantum number this shift is mainly determined by the Coulomb dipole exchange interaction and can be approximated by the expression:
\begin{eqnarray}
\label{eq9}
\fl \delta \overline{E}(K_0 nl^{4l+1}n(l+1)^{N_2+1}-K_0 nl^{4l+2}n(l+1)^{N_2}) \nonumber \\
\lo{\simeq} \frac{(4l+5-N_2)(l+1)}{4l+5} \left[ \frac{2}{3}-\frac{1}{2(2l+1)(2l+3)} \right] G^1_K(nl,n(l+1))
\end{eqnarray}
where $K_0$ means the closed "passive" shells and $G^1_K$ is the Slater exchange integral for the initial configuration $K$. Thus the shift $\delta \overline{E}$ obtains the largest positive value at the smallest number of electrons in the open shell and decreases with increasing $N_2$. This tendency is illustrated by the variation of spectra $3{\rm p}^53{\rm d}^N \rightarrow 3{\rm p}^63{\rm d}^{N-1}$ in the isonuclear sequence for the ions of Co (figure \ref{allCo}). Upon filling the outer open $3{\rm d}^N$ shell the concentration of intensive lines on the high-energy side gradually diminishes and they spread within a larger part of the interval of possible transition energies. The similar variation of the spectra arising from the open $4{\rm d}$ and $4{\rm p}$ shells takes place for the Sm isonuclear sequence \cite{O'Sullivan}. Consideration of the average energy of Auger transitions (\ref{eq7}) shows that the enhancement of high-energy lines and suppression of low-energy lines are most pronounced at a small number of electrons in the outer shell \cite{JKK97}. 

	Therefore the formulated selection rule with respect to the number of vacancy-electron pairs can have a fairly  wide application in the interpretation of various spectra.

\end{document}